\documentclass[preprint,floats,aps]{revtex4}
\usepackage{dcolumn}
\usepackage{graphics}
\usepackage{rotating}
\usepackage{xspace}
\usepackage{epsfig}
\usepackage{subfigure}     		
\usepackage{latexsym}
\usepackage{amsfonts}

\begin{document}

\preprint{\vbox{\hbox {January 2006} }}

\title{Conformal Transformations and Accelerated Cosmologies} 
\author{\bf James L. Crooks and Paul H. Frampton}
\affiliation{University of North Carolina, Chapel Hill, NC  27599-3255, USA.}

\date{\today}

\begin{abstract}
A cosmological theory that predicts a late-time accelerated attractor with
a constant dark matter to dark energy ratio can be said to solve the 
Coincidence
Problem.  Such cosmologies are naturally generated in the context of 
non-standard gravity theories under conformal transformation
because of the resulting couplings between scalar fields 
and matter.  The present work examines 
four classes of these transformed theories and finds that only a small 
subset--those with a single scalar field--are capable 
of solving the Coincidence Problem. 
\end{abstract}
\maketitle

\newpage

\section{Introduction}
\label{sec:outline}

The cosmological Coincidence Problem stems from the observation that the 
present-day ratio of dark matter (or total matter) to dark energy is near 
unity despite the fact that these two species are thought to evolve 
with time in different ways.  Perhaps it is just a coincidence; 
however, physicists are weary of accepting it as such without 
investigating alternative explanations.  One possibility is that the two
species do not actually evolve indepentently.  Indeed, a coupling 
between dark matter and 
dark energy can naturally yield a constant energy ratio near unity
in the asymptotic time limit.  If this 
steady-state is allowed and found to be consistent with a 
accelerated and dynamically stable universe then
we can say that the theory generating it solves the Coincidence
Problem. 

Toward this end, a number of coupling schemes have been put forward
({\it e.g.} \cite{zlat99} and \cite{zimd01a}).  We will focus here on 
the scenario which involves conformally 
transforming a non-standard gravity theory.  
Generally, the alternative theory of gravity is taken to be the Brans-Dicke 
theory \cite{bran61} (the simplest and best-motivated extension
of general relativity relativity), in which 
Newton's constant is re-interpreted as a time-varying scalar field 
\cite{damo90}.  
Amendola and various co-authors \cite{tocc02} \cite{amen01a}, as well 
as other groups
(\textit{e.g.} \cite{bean01a} and \cite{torr02}), have shown that the 
coupling derived from a Brans-Dicke 
theory yields a stable and accelerated critical point on which there is
a constant ratio of dark matter to dark energy, 
thus solving the Coincidence Problem.  However, many other 
gravity theories can
be conformally transformed with differing results for the ``Einstein frame'' 
theory (see \cite{fara99} for a review).    

This article examines four classes of generalized gravity 
theories to find those that are incapable of solving the Coincidence Problem.
It is shown that most kinds of non-standard gravity theories do not, upon 
conformal transformation, 
contain the accelerated, stable, and mixed critical points that are required.  
Lagrangians involving scalar fields coupled to terms quadratic in the Ricci 
scalar, 
as well as theories involving higher derivatives of the Ricci scalar, fail 
to solve the Coincidence Problem.  On the other hand, both 
scalar-tensor theories linear in the Ricci scalar
(like the Brans-Dicke theory) and nonlinear theories without scalar fields 
can contain the 
required critical points but only in special cases. The general requirements 
are that upon conformal transformation the theory must contain only 
one scalar field, and that
the dark matter-dark energy coupling function $\Upsilon(\phi)$ must be 
proportional to the square root of the scalar 
field coupling to gravity $\omega(\phi)$.

Section \ref{sec:contrangct} presents the method for conformally 
transforming a non-standard theory of gravitation.  
Sections \ref{sec:contranst} through
\ref{sec:contranho} illustrate the practical use of conformal 
transformations by applying them to theories of 
four general types:
linear scalar-tensor, nonlinear, quadratic scalar-tensor, 
and higher-order theories.  
The physical meaning of the conformal 
transformation is discussed in Section \ref{sec:contranint}.  

A general consequence of conformal
transformations is the appearance of a new coupling between a scalar field 
and the matter fields such that the stress-energy
tensor of the matter fields is no longer conserved.  Section 
\ref{sec:contrancos} describes how the scalar
field may be identified with dark energy.  Particle physics experiments and 
astrophysical
observations put severe limits on the non-conservation of normal 
baryonic matter, so Section 
\ref{sec:contrandmde} describes various ways to avoid these constraints.  One 
way, devised by Damour
\textit{et al.} \cite{damo90}, is to couple the scalar field only to dark 
matter with an interaction mediated by the
time-derivative of the scalar field multiplied by the energy density of dark
 matter.
The modern version of this model is called the Coupled Quintessence 
Model.  An alternative way introduced here 
also involves coupling a scalar field to dark matter, but 
now the interaction is mediated by baryons and 
non-relativistic neutrinos as well as well as dark matter itself.

Finally, to solve the Coincidence Problem we must 
determine which kinds of couplings
allow the universe to converge to an accelerated, stable state 
containing a constant
ratio of dark matter to dark energy.  The first and last of these 
requirements reduce the
space of possible schemes enormously.  Sections \ref{sec:cosdynasymsingle} 
and 
\ref{sec:cosdynasymdouble} explain how a single scalar field model can meet 
these two requirements but 
a dual scalar field model cannot.  Stability will not be discussed here
as it is more dependent on the particular choices made for various
arbitrary functions and parameter values.  
The general argument against two or more scalar fields is 
given in Section \ref{sec:cosdynasymgeneral} along with an explanation of why
the single scalar field case is different.  This section also
shows that for a single scalar field to work there 
are constraints on the possible choices of functions
for the scalar coupling to gravity, the dark matter-dark energy coupling, and 
the scalar field potential.

\section{General Conformal Transformations}
\label{sec:contrangct}
From a general action in a scalar-tensor or nonlinear gravity theory,

\begin{equation}\label{JordanFrameS}
\widetilde{S} = \int d^4 x \left\{ \sqrt{-\widetilde{g}} \left[ F(\psi,
\widetilde{R}) 
- A(\psi) \partial_\mu \psi \partial_\nu \psi 
\widetilde{g}^{\mu \nu}\right] -  \widetilde{L}^{(M)}\right\}, 
\end{equation}

\noindent
it is possible to generate the analogue of the Einstein field equations:

\begin{eqnarray} \label{jordframact}
0 &=& \left( \frac{\partial F(\psi,\widetilde{R})}{\partial \widetilde{R}} 
\right) 
(\widetilde{R}_{\mu \nu} - \frac{1}{2} \widetilde{g}_{\mu \nu} \widetilde{R}) 
+ \frac{1}{2} \widetilde{g}_{\mu \nu} \left( \frac{\partial F(\psi,\tilde{R})
}
{\partial \widetilde{R}} \widetilde{R} -F(\psi,\widetilde{R})\right) 
\nonumber\\
& & {} - A(\psi) \left( \partial_\mu \psi \partial_\nu \psi - \frac{1}{2} 
\widetilde{g}_{\mu \nu} \widetilde{g}^{\rho \sigma} \partial_\rho \psi 
\partial_\sigma 
\psi \right) + \widetilde{g}_{\mu \nu} \widetilde{\square} \left( \frac{
\partial 
F(\psi,\widetilde{R})}{\partial \widetilde{R}} \right) \nonumber\\
& & {} - \widetilde{\nabla}_{\mu} \widetilde{\nabla}_{\nu}
\left( \frac{\partial F(\psi,\widetilde{R})}{\partial \widetilde{R}}  
\right) - \frac{1}{2} \widetilde{T}_{\mu \nu}^{(M)}.
\end{eqnarray}

\noindent
The two terms involving second derivatives of $\partial 
F(\psi,\widetilde{R})/\partial \widetilde{R}$ are due to the non-vanishing 
variation
of the Riemann tensor in this general situation.  Their derivation 
is given in \cite{adle65}.

Conformal transformations are transformations of 
the metric 
$\widetilde{g}_{\mu \nu}$ such that a nonstandard gravity theory may
be written in a new form called the 
Einstein frame, the "frame" where the gravitational part of the action 
looks like 
the normal Einsteinian one (see \cite{maed89} for the original discussion of 
this general case).  

To begin, define the conformal function $\Omega^2$ in terms of the Jordan 
frame gravitational
action:

\begin{equation} \label{omegadef}
\left| \frac{\partial F(\psi,\widetilde{R})}{\partial \widetilde{R}} \right| 
\equiv \frac{\Omega^2}{16 \pi G}.
\end{equation}

\noindent
The conformal transformation (not a coordinate transformation in the usual 
sense) is 
then defined by the following transformation of the metric: 

\begin{eqnarray} 
g_{\mu \nu} &=& \Omega^2 \widetilde{g}_{\mu \nu} \\
g^{\mu \nu} &=& \Omega^{-2} \widetilde{g}^{\mu \nu} \\
\sqrt{-g} &=& \Omega^4 \sqrt{-\widetilde{g}}.
\end{eqnarray}

\noindent
Because the gravitational part of the action is defined by the metric and 
its derivatives 
the geometrical side of the Einstein equations must transform as well: 

\begin{eqnarray} \label{riccitrans}
R_{\mu \nu} - \frac{1}{2} g_{\mu \nu} R &=& \widetilde{R}_{\mu \nu} 
- \frac{1}{2} \widetilde{g}_{\mu \nu} \widetilde{R} - \Omega^{-2} \widetilde{
\nabla}_\mu 
\widetilde{\nabla}_\nu \Omega^2 + \Omega^{-2} \widetilde{g}_{\mu \nu} 
\widetilde{\square} 
\Omega^2 \nonumber\\ 
& & {} + \frac{3}{2} \Omega^{-4} \left( 
\partial_\mu \Omega^2 \partial_\nu \Omega^2 - \frac{1}{2} \widetilde{g}_{\mu 
\nu}
\widetilde{g}^{\alpha \beta} \partial_\alpha \Omega^2 \partial_\beta 
\Omega^2 \right).
\end{eqnarray}
 
\noindent
The extra terms thus produced can be used to cancel the extra terms in 
equation (\ref{jordframact}).
Let $\zeta = \left( \partial F(\psi,\widetilde{R})/\partial \widetilde{R} 
\right) / \left| 
\partial F(\psi,\widetilde{R})/\partial \widetilde{R} \right|$.  
Then, with the above substitution, equation (\ref{jordframact}) reads

\begin{eqnarray}\label{einfreineq}
R_{\mu \nu} - \frac{1}{2} g_{\mu \nu} R &=& \zeta \Bigg[ 8 \pi G \Omega^{-4} 
g_{\mu \nu} 
\left( F(\psi,\tilde{R}) - \frac{\partial F(\psi,\widetilde{R})}{\partial 
\widetilde{R}} 
\widetilde{R} \right) + 8 \pi G \Omega^{-2} 
\widetilde{T}_{\mu \nu}^{(M)}  \nonumber\\
& & {}  + 16 \pi G \Omega^{-2} A(\psi) \left( \partial_{\mu} \psi \partial_{
\nu} \psi - 
\frac{1}{2} g_{\mu \nu} g^{\rho \sigma} \partial_{\rho} \psi \partial_{
\sigma} \psi \right) \Bigg] \nonumber\\
& & {} + \frac{3}{2} \Omega^{-4} \left( \partial_{\mu} \Omega^2 \partial_{
\nu} \Omega^2 
- \frac{1}{2} g_{\mu \nu} g^{\alpha \beta} \partial_{\alpha} \Omega^2 
\partial_{\beta} \Omega^2 \right).
\end{eqnarray}

In those cases where equation (\ref{omegadef}) may be written in the form 
$\widetilde{R} = 
\widetilde{R}(\Omega^2,\psi)$ it is possible to rewrite the second term on 
the right hand
side of equation (\ref{einfreineq}) as a scalar potential,

\begin{equation}
U(\psi,\Omega^2) \equiv \Omega^{-4} \left[ \zeta F(\psi,\widetilde{R}) - 
\left| \frac{\partial 
F(\psi,\widetilde{R})}{\partial \widetilde{R}} \right| \widetilde{R} \right] 
\Bigg|_{\widetilde{R} = 
\widetilde{R}(\Omega^2,\psi)}. 
\end{equation}

\noindent 
It is also possible to rewrite this term when equation (\ref{omegadef}) does 
not involve 
$\widetilde{R}$, i.e. when $F(\widetilde{R},\psi)$ is linear in $\widetilde{
R}$.  
Then $U(\psi,\Omega^2)$ is simply proportional to the scalar 
potential term in 
$F(\widetilde{R},\psi)$ if such a term exists, and zero otherwise. 

Since the Einstein equations should by definition have the normal form in the 
Einstein frame,
equation (\ref{einfreineq}) implies

\begin{eqnarray} \label{sclstrsenrgy}
T_{\mu \nu}^{(\psi,\Omega^2)} &\equiv& 2 \zeta A(\psi)  \Omega^{-2} \left( 
\partial_{\mu} 
\psi \partial_{\nu} \psi - \frac{1}{2} g_{\mu \nu} g^{\rho \sigma} \partial_{
\rho} \psi 
\partial_{\sigma} \psi \right) + g_{\mu \nu} U(\psi,\Omega^2) \nonumber\\
& & {} + \frac{3}{16 \pi G} \Omega^{-4} \left(\partial_{\mu} \Omega^2 
\partial_{\nu} \Omega^2 
- \frac{1}{2} g_{\mu \nu} g^{\alpha \beta} \partial_{\alpha} \Omega^2 
\partial_{\beta} \Omega^2 \right) 
\end{eqnarray}

\noindent 
and

\begin{equation}
T_{\mu \nu}^{(M)} \equiv \zeta \Omega^{-2} \widetilde{T}_{\mu \nu}^{(M)}.
\end{equation}

\noindent 
With these definitions equation (\ref{einfreineq}) then reads

\begin{equation} \label{einseq}
R_{\mu \nu} - \frac{1}{2} g_{\mu \nu} R = 8 \pi G \left( T_{\mu \nu}^{(M)} 
+ T_{\mu \nu}^{(\psi,\Omega^2)}\right).
\end{equation}

\noindent 
However, since the relation

\begin{equation}
\widehat{T}_{\mu \nu} = \frac{2}{\sqrt{-\widehat{g}}} \frac{\delta 
\widehat{L}}
{\delta \widehat{g}^{\mu \nu}}
\end{equation}

\noindent 
must be valid in either frame, we can reverse-engineer equation 
(\ref{sclstrsenrgy}) 
to get an expression for the Lagrangian in the Einstein frame,

\begin{eqnarray} \label{einslagr}
L^{(\psi,\Omega^2)} &=& \sqrt{-g} \bigg[ \frac{}{} \zeta A(\psi) \Omega^{-2} 
g^{\mu \nu} 
\partial_{\mu} \psi \partial_{\nu} \psi - U(\psi,\Omega^2) \nonumber\\
& &  {} + \frac{3}{32 \pi G } \Omega^{-4} g^{\mu \nu} \partial_{\mu} 
\Omega^2 \partial_{\nu} \Omega^2 \bigg].
\end{eqnarray}

\noindent
Then with this identity we can write down the Einstein frame action that 
generates equation (\ref{einseq}),

\begin{equation} \label{einsact}
S = \int d^4 x \left\{ \frac{\sqrt{-g} R}{16 \pi G} - L^{(\psi,\Omega^2)} - 
L^{(M)} \right\}.
\end{equation}

An important property of matter fields under conformal transformation is that 
a 
stress-energy tensor conserved in the Jordan frame is not generally conserved
in the Einstein frame and vice-versa.  To see this, first expand the
4-divergence
of the stress-energy tensor of matter species X in the Jordan frame and then 
substitute appropriate factors of $\Omega^2$ \cite{carm82}:

\begin{eqnarray} \label{noncons}
\widetilde{\nabla}_{\mu} \widetilde{T}^{\mu (X)}_{\nu} &=& \frac{1}{\sqrt{- 
\widetilde{g}}}
\partial_{\mu} \left( \sqrt{- \widetilde{g}} \widetilde{T}^{\mu (X)}_{\nu} 
\right) - \frac{1}{2}
\widetilde{T}^{\mu \gamma (X)} \partial_{\nu} \widetilde{g}_{\mu \gamma} 
\nonumber\\
&=& \frac{\Omega^4}{\sqrt{-g}} \partial_{\mu} \left( \sqrt{-g} T^{\mu (X)}_{
\nu}
\right) - \frac{1}{2} \Omega^6 T^{\mu \gamma (X)} \partial_{\nu} \left( 
\Omega^{-2} g_{\mu \gamma}
 \right) \nonumber\\
 &=& \Omega^4 \nabla_{\mu} T^{\mu (X)}_{\nu} + \frac{1}{2} \Omega^2 \partial_{
\nu} \Omega^2 T^{(X)} \nonumber\\
 \nabla_{\mu} T^{\mu (X)}_{\nu} &=& \Omega^{-4} \widetilde{\nabla}_{\nu} 
\widetilde{T}^{\mu (X)}_{\nu}
 - \frac{1}{2} \Omega^{-6} \partial_{\nu} \Omega^2 \widetilde{T}^{(X)}
\end{eqnarray} 

\noindent 
Therefore, when the stress-energy tensor is conserved in one frame it is not 
conserved in 
the other except if the species X is radiation 
($\widehat{T}^{(R)} = 0$) or $\Omega^2$ is 
constant.  

To see this another way, vary the action to get the equations of motion in 
the Einstein frame.  
Varying equation (\ref{einsact}) with respect to $\psi$ generates the $\psi$ 
equation of motion

\begin{eqnarray}
2 \zeta A(\psi) \Omega^{-2} \square \psi &=& 2 \zeta A(\psi) \Omega^{-4} g^{
\mu \nu} \partial_{\mu}
\Omega^2 \partial_{\nu} \psi - \zeta \frac{\partial A(\psi)}{\partial \psi} 
\Omega^{-2} 
g^{\mu \nu} \partial_{\mu} \psi \partial_{\nu} \psi \nonumber\\
& & - \frac{\partial U(\psi,\Omega^2)}{\partial \psi}. 
\end{eqnarray}

\noindent
Varying the action with respect to $\Omega^2$ (using the trace of equation 
(\ref{riccitrans}) to transform between frames) yields the equation of motion 
for $\Omega^2$.  
Because of the transformation between frames this equation includes a Ricci 
scalar term which
may be eliminated by substituting the trace of equation (\ref{einfreineq}).  
Doing so yields

\begin{eqnarray}
\square \Omega^2 &=& \Omega^{-2} g^{\mu \nu} \partial_{\mu} \Omega^2 
\partial_{\nu} \Omega^2 
- \frac{16 \pi G}{3} \zeta A(\psi) g^{\mu \nu} \partial_{\mu} \psi \partial_{
\nu} \psi \nonumber\\
& & - \frac{16 \pi G}{3} \Omega^4 \frac{\partial U(\psi,\Omega^2)}{\partial 
\Omega^2}
+ \frac{8 \pi G}{3} \Omega^2 T^{(M)}.
\end{eqnarray} 

\noindent
Finally, we take the 4-divergence of equation (\ref{einseq}) and substitute in 
the equations of motion.  
After cancellation the only terms remaining give

\begin{equation}\label{matcons}
\nabla^{\mu} T_{\mu \nu}^{(M)} = - \frac{1}{2} \Omega^{-2} \partial_{\nu} 
\Omega^2 T^{(M)}.
\end{equation} 

However, since the 4-divergence must evaluate to zero on both sides of the 
Einstein equations
(equation (\ref{einfreineq})) we can read off:

\begin{equation}\label{secons}
\nabla^{\mu} T_{\mu \nu}^{(\psi,\Omega^2)} = \frac{1}{2} \Omega^{-2} 
\partial_{\nu} \Omega^2 T^{(M)}.
\end{equation}

\noindent
Note that this does not contradict equation (\ref{noncons}) since $\psi$ and 
$\Omega^2$ are not
``matter fields'' in the sense used above, \textit{i.e.}, they do not appear 
in the matter part of the 
original action.

\section{Interpretation}
\label{sec:contranint}

What is the physical significance of a conformal transformation?  
What is the relationship between the Jordan frame and 
the Einstein frame?
If we assume that only the Einstein frame is directly observable, is the 
Jordan frame
anything more than a mathematical device?  There is much debate surrounding 
these questions (see, \textit{e.g.}, \cite{magn94}, \cite{gott90}, and 
\cite{barr88}). 
Different authors interpret conformal transformations in different ways; the 
debate often
revolves around the question: which of the frames is ``physical''?

Some simply define the ``physical'' frame as the one in which the matter 
stress-energy 
tensor is conserved.  Others define it as the one in which we happen to be 
making 
observations, which is not necessarily the same thing.  Neither position can 
be rejected out of hand.  It seems 
that either frame may be pressed into service as the ``physical'' frame 
depending on the use 
to which the theory is being put.  Furthermore, even once the ``physical'' 
frame is chosen, 
the meaning of the other frame is not obvious.  
Is it just a convenient fiction?  Is it 
possible for both 
frames to be physically meaningful? 

Perhaps the most galling problem with conformal transformations theories is 
that often no interpretation--or even agnosticism--is stated explicitly.  
The interpretation given
in the present work depends on the way in which the interaction term is 
partitioned among the different 
matter species, as will be discussed in Section \ref{sec:contrandmde}.  For 
now we will dodge the question and look at various
classes of non-standard gravity theories and what happens to them upon 
conformal transformation.

\section{Linear Scalar-Tensor Theories}
\label{sec:contranst}

Linear scalar-tensor theories are those for which 
$F(\psi,\widetilde{R})$ in equation 
(\ref{JordanFrameS}) takes the form $F(\psi,\widetilde{R}) = f(\psi) 
\widetilde{R} 
+ V(\psi)$ \cite{berk91}.  In this case $\zeta$ takes the form $\zeta = 
|f(\psi)|/f(\psi)$.  

Since these theories are linear in $\widetilde{R}$ it is obvious that 
$\Omega^2$ is a function only of 
$\psi$.  This has two consequences.  First, if $V(\psi)$ exists then

\begin{equation}
U(\psi,\Omega^2) = V(\psi) \frac{|f(\psi)|}{f^3(\psi)} \mbox{ when $f(\psi) 
\neq$ 0},
\end{equation}

\noindent
though presumably $V(\psi)$ could be chosen such that the ratio on the right 
hand side is defined when
$f(\psi) = 0$.  
Second, there is effectively only a single scalar field in the action since 
the two kinetic 
terms can be directly added together.  Going back to equation 
(\ref{einslagr}) we can therefore 
make the substitution

\begin{equation} \label{phidef}
\phi = \int d \psi\sqrt{\frac{3}{16 \pi G} \Omega^{-4}(\psi) \left( \frac{
\partial \Omega^2(\psi)}{\partial \psi}
 \right)^2 + 2 \zeta A(\psi)\Omega^{-2}(\psi)} 
\end{equation}

\noindent
to get a single minimally coupled scalar field in the action.  Now instead 
of two equations of 
motion there is only one,

\begin{equation}\label{regboxeqn}
\square \phi + \frac{\partial U(\phi)}{\partial \phi} = \frac{1}{2} 
\Omega^{-2}(\phi) \frac{d 
\Omega^2(\phi)}{d \phi }T^{(M)}.
\end{equation}

\noindent
Of course we could also set up equation (\ref{phidef}) such that the 
resulting action would contain a 
non-minimally coupled scalar field; the choice is arbitrary, though the
minimally coupled option is favored on aesthetic grounds. 

Equation (\ref{phidef}) also implies that when $f(\psi) \propto \psi^n$ and
$A(\psi) \propto \psi^{n - 2}$ for a given $n$ the resulting Einstein frame 
equations of motion are 
indistinguishable from the Brans-Dicke ($n=1$) case.  The same applies if 
both $f(\psi)$ and $A(\psi)$ are proportional
to $e^{\mu \psi}$, or in any other circumstance in which $A(\psi) \propto 
\frac{1}{f(\psi)} 
\left(\frac{\partial f(\psi)}{\partial \psi}\right)^2$ holds.
Curiously, this indistinguishability means that if the Jordan frame were 
inferred to be the natural one for the dark sector there would be no way 
to distinguish one such theory
from any other since they are all equivalent in the Einstein 
frame.

\section{Nonlinear Gravity Theories}
\label{sec:contranng}

Nonlinear gravity theories are those for which $F(\psi,\widetilde{R})$ do 
not depend on $\psi$, 
\textit{i.e.}, $F(\psi,\widetilde{R})=F(\widetilde{R})$ \cite{barr88}.  Since 
there are no scalar fields coupled 
to gravity in the Jordan frame we can ignore $A(\psi)$ and $V(\psi)$.  This 
means that the 
resulting Einstein frame potential depends entirely on the Jordan frame 
theory chosen--there is 
no freedom to choose the functional form.  The Einstein frame equations of 
motion, though, are of exactly
the same form as those in the linear scalar-tensor case given by equation 
(\ref{regboxeqn}) once $\Omega^2$ has been transformed into the minimally 
coupled field $\phi$ using 

\begin{equation}\label{omtophi}
\Omega^2 = e^{\sqrt{\frac{16 \pi G}{3}} \phi}.
\end{equation}

\noindent
However,
because $\Omega^2$ can often be defined as a function of $\widetilde{R}$ it 
is sometimes possible to 
relate $\widetilde{R}$ directly to $\phi$.  The absolute value in equation 
(\ref{omegadef})
complicates the matter since two values of $\widetilde{R}$ can correspond to 
any one value of $\Omega^2$.  
In practice, however, one of the two $\zeta$ values may not be available 
since $F(\widetilde{R})$ 
may be assumed real and $F'(\widetilde{R})$ is sometimes intrinsically 
positive or negative.  
On the other hand, even if both values are available we are effectively free 
to choose which one
we want since the value of $\widetilde{R}$ is not observable anyway.

In addition to $\zeta$, however, there is another sign ambiguity when 
$F'(\widetilde{R})$ is an even 
function of $\widetilde{R}$.  For 
example when f $F(\widetilde{R}) \propto \widetilde{R} (1+(\alpha 
\widetilde{R})^{-2})$ there 
are four values of $\widetilde{R}$ corresponding to each negative value of 
$\phi$; specifically, there 
is a splitting due to $\zeta$ and another due to the symmetry of $F'(
\widetilde{R})$, for a total of 
four solutions.  

It is interesting to compare these theories 
to the cubic one: ($F(\widetilde{R}) \propto \widetilde{R} + 
\alpha \widetilde{R}^2 + 
\gamma \widetilde{R}^3$).  When $\alpha^2/\beta > 3$ the cubic theory also 
has four possible values of 
$\widetilde{R}$ for certain values of $\phi$.  However, they are not 
symmetrically distributed and 
therefore the corresponding potentials cannot in general
be written in a simple form involving $\zeta$ and $\pm$.  
The problem of having too many Einstein-frame potentials obviously gets worse 
with higher-order polynomials.
  
Also, despite the fact that each Jordan frame theory corresponds to a unique 
potential (modulo possible sign ambiguities) in the Einstein frame, 
there are theories outside this class (\textit{e.g.}, linear 
scalar-tensor theories) whose equation of motion is 
the same and whose potentials are completely arbitrary.  Thus there is no way
to distinguish any Jordan-frame 
theory in this class from a scalar-tensor theory using only the 
Einstein-frame potential.

\section{Quadratic Scalar-Tensor}
\label{sec:contranhyb}

Quadratic scalar-tensor theories are those for which $F(\psi,\widetilde{R})$ 
takes the form

\begin{equation}\label{Fdefhyb}
F(\psi,\widetilde{R}) = f(\psi) \widetilde{R} + g(\psi) \widetilde{R}^2 
+ V(\psi).
\end{equation}

\noindent
The two best known models in this class are the dilaton \cite{maed88} 
and renormalizable \cite{futa89} theories.  
Unlike the nonlinear gravity and linear scalar-tensor cases, 
$\Omega^2$ here involves both a function of $\psi$ and 
a nonlinear power of $\widetilde{R}$.  This means we must keep 
both kinetic terms in equation (\ref{sclstrsenrgy}); we can neither combine 
them as in the nonlinear 
gravity examples nor eliminate one as in the linear scalar-tensor examples.  
However, we can transform $\Omega^2$ 
into $\phi$ as before to make at least one of the kinetic terms minimally 
coupled:

\begin{eqnarray}\label{hybtmunu}
T_{\mu \nu}^{(\psi,\phi)} &\equiv& 2 \zeta A(\psi)  e^{-\sqrt{\frac{16 \pi G}
	{3}} \phi} 
	\left( \partial_{\mu} \psi \partial_{\nu} \psi - \frac{1}{2} g_{\mu 
	\nu} g^{\rho \sigma} 
	\partial_{\rho} \psi \partial_{\sigma} \psi \right) + g_{\mu \nu} 
	U(\psi,\phi) \nonumber\\
& & {} + \partial_{\mu} \phi \partial_{\nu} \phi - \frac{1}{2} g_{\mu \nu} 
	g^{\alpha \beta} 
	\partial_{\alpha} \phi \partial_{\beta} \phi.  
\end{eqnarray}   

\noindent
This leads to a potential,

\begin{equation}
U(\psi,\phi) = \zeta V(\psi) e^{-2 \sqrt{\frac{16 \pi G}{3}} \phi} - \frac{
	\zeta}{4 g(\psi)} \left[ 
	\zeta f(\psi) e^{-\sqrt{\frac{16 \pi G}{3}} \phi} - \frac{1}{16 \pi 
	G}\right]^2,
\end{equation}

\noindent
an equation of motion for $\psi$,

\begin{eqnarray}\label{boxpsi}
\square \psi + \frac{\zeta}{2A(\psi)}e^{\sqrt{\frac{16 \pi G}{3}} \phi} 
\frac{\partial U(\psi,\phi)}
{\partial \psi} &=& -\frac{1}{2} \frac{\partial \log A(\psi)}{\partial \psi} 
e^{-\sqrt{\frac{16 \pi G}
{3}} \phi} g^{\mu \nu}\partial_{\mu} \psi \partial_{\nu} \psi \nonumber\\& & 
{}+ \sqrt{\frac{16 \pi G}{3}} 
g^{\mu \nu} \partial_{\mu} \psi \partial_{\nu} \phi,  
\end{eqnarray}

\noindent
and an equation of motion for $\phi$,

\begin{eqnarray}\label{boxphi}
\square \phi + \frac{\partial U(\psi,\phi)}{\partial 
\phi} &=& -\zeta \sqrt{\frac{16 \pi G}{3}} A(\psi)e^{-\sqrt{\frac{16 \pi G}
{3}} \phi} g^{\mu \nu} 
\partial_{\mu} \psi \partial_{\nu} \psi + \frac{1}{2} \sqrt{\frac{16 \pi G}
{3}} T^{(M)}. \nonumber\\ & &
\end{eqnarray}

These theories do not share the distinguishability problem of the previous 
two types since there are
two coupled scalar fields with non-trivial equations of motion and a 
potential determined uniquely 
(modulo $\zeta$) by the specific Jordan-frame theory.  It would be quite 
bizarre for arbitrary
functions in some other Jordan-frame theory to reproduce these equations.

\section{Higher-Order Gravity Theories}
\label{sec:contranho}

There are many gravitational actions without scalar fields we have thus far 
neglected.  For example we have looked 
only at theories involving the Ricci scalar and have ignored the Ricci and 
Riemann tensors.  It is possible to 
imagine theories with terms in the action proportional to $\widetilde{R}_{\mu 
\nu \rho \sigma} 
\widetilde{R}^{\mu \nu \rho \sigma}$ or to $\widetilde{R}_{\mu \nu} 
\widetilde{R}^{\mu \nu}$.  However, 
the former may be eliminated as part of a total divergence and the latter is
reducible to a form involving only $\widetilde{R}^2$ in a 
Friedmann-Robertson-Walker metric \cite{wand94} 
as will be assumed throughout this paper.

Alternatively, we could look at actions involving functions of derivatives of 
the Ricci scalar: $F(\widetilde{R},
\widetilde{\square} \widetilde{R},\dots,\widetilde{\square}^{n} 
\widetilde{R})$ (see \cite{wand94}, \cite{schm90}, and \cite{gott90}).
Wands \cite{wand94} has shown that, by a redefinition of variables (not, it 
should be emphasized, 
a conformal transformation) $\psi_{i} \equiv
\widetilde{\square}^{i} \widetilde{R}$, we may write this general action as

\begin{equation}\label{highgravact}
\widetilde{S} = \int d^{4}x \left\{ \sqrt{-\widetilde{g}}\left[\left( 
\sum^{n}_{j=0} \widetilde{\square}^{j}
\frac{\partial F}{\partial \psi_{j}}\right)\widetilde{R} + F -\sum^{n}_{j=0}
\psi_{j}
\frac{\partial F}{\partial \psi_{j}}\right]-\widetilde{L}^{(M)}\right\} 
\end{equation}

\noindent
where $F$ is now written as $F(\psi_{0},\dots,\psi_{n})$.  Thus a 
higher-order gravity theory may be re-written 
in a form
similar to that of a linear or quadratic scalar-tensor theory.  
In fact, this similarity is quite deep.  Wands gives the
example $F(\widetilde{R},\widetilde{\square} \widetilde{R}) \propto 
\widetilde{R}+\gamma \widetilde{R}\widetilde{\square} 
\widetilde{R}$ and shows that under conformal transformation it has a 
Lagrangian
of the form given by equation (\ref{einslagr}).  Redefining $\Omega^2 = 
\exp(\sqrt{\frac{16 \pi G}{3}}\phi)$
and $\psi_{0} = \sqrt{\frac{8 \pi G}{\gamma}}\psi$ the stress-energy tensor 
is then just equation (\ref{hybtmunu}) 
with $\zeta A(\psi)=\frac{1}{2}$.  Furthermore, the form of the potential 
is uniquely determined to be

\begin{equation}
U(\psi,\phi) = \frac{\psi}{\sqrt{32 \pi G \gamma}} \left(e^{-\sqrt{\frac{16 
\pi G}{3}}\phi} - 
e^{2\sqrt{\frac{16 \pi G}{3}}\phi}\right).
\end{equation}

\noindent
Where does the kinetic term for $\psi_{0}$ (or equivalently $\psi$) come 
from?  When we replace $\psi_{1}$ with 
$\Omega^2$ a new term proportional to $\psi_{0} \widetilde{\square} 
\psi_{0}$ appears in the action, which, after integration
 by parts, becomes a kinetic term for $\psi_{0}$.  

What about the general case?  If we take the conformal function to be

\begin{equation}
\Omega^2= \sum^{n}_{j=0}\widetilde{\square}^{j}\frac{\partial F}{\partial 
\psi_{j}}
\end{equation}

\noindent
and convert to the Einstein frame the result depends strongly on the number 
of derivatives of the Ricci scalar.
Quadratic or higher polynomials in $\widetilde{\square} \widetilde{R}$ yield 
three or more scalar fields, while terms of order 
$\widetilde{\square}^2 \widetilde{R}$, $\widetilde{\square}^3 \widetilde{R}$,
etc..., yield more than two scalar 
fields and/or kinetic terms involving higher 
derivatives.  The most general form of higher-order gravity theory that has 
an Einstein frame Lagrangian 
equivalent to equation (\ref{einslagr}) appears to be

\begin{equation}
F(\widetilde{R},\widetilde{\square} \widetilde{R}) = f_{0}(\widetilde{R})
+f_{1}(\widetilde{R})\widetilde{\square} \widetilde{R}.
\end{equation} 

\noindent
The Einstein frame version of the theory has the potential

\begin{equation}
U(\psi_{0},\Omega^2) = \Omega^{-4} \left(f_{1}(\psi_{0})-\Omega^2 \psi_{0} 
\right),
\end{equation}

\noindent
and we may identify

\begin{equation}
\zeta A(\psi_{0}) = \frac{\partial f_{1}(\psi_{0})}{\partial \psi_{0}}.
\end{equation}

By contrast, a theory linear in $\widetilde{\square}^{2}\widetilde{R}$ 
instead of just $\widetilde{\square}\widetilde{R}$,

\begin{equation}
F(\widetilde{R},\widetilde{\square}\widetilde{R},\widetilde{\square}^{2} 
\widetilde{R}) = f_{0}(\widetilde{R})
+f_{1}(\widetilde{R})g(\widetilde{\square} \widetilde{R})+f_{2}(
\widetilde{R})\widetilde{\square}^{2}\widetilde{R},
\end{equation}

\noindent
cannot be written in a form equivalent to equation (\ref{einslagr}).  In 
this case we find that the second and third
terms under the brackets in equation (\ref{highgravact}), 
after replacing $\psi_{2}$ with $\Omega^2$,
may be written as

\begin{eqnarray}
F - \sum_{j=0}^{2}\psi_{j}\frac{\partial F}{\partial \psi_{j}} &=& f_{0}
(\psi_{0})+f_{1}(\psi_{0}) 
g(\psi_{1})-f_{1}(\psi_{0})\frac{d g(\psi_{1})}{d \psi_{1}}\psi_{1}+\frac{
d f_{2}(\psi_{0})}{d \psi_{0}}\psi_{1}^{2}
\nonumber\\ & &-\psi_{0} \Omega^{2} + \left(\frac{d^{2} f_{2}(\psi_{0})}{
d \psi_{0}^{2}}\psi_{1} - 
\frac{d f_{1}(\psi_{0})}{d \psi_{0}}\frac{d g(\psi_{1})}{d \psi_{1}}\right) 
\widetilde{g}^{\mu \nu}
\partial_{\mu}\psi_{0} \partial_{\nu}\psi_{0} \nonumber\\ & &-f_{1}(\psi_{0})
\frac{d^{2} g(\psi_{1})}{d \psi_{1}^{2}} \widetilde{g}^{\mu \nu} \partial_{
\mu}\psi_{0} \partial_{\nu}\psi_{1}.
\end{eqnarray}

There are now three different kinetic terms in the total Einstein-frame 
action (the third being the one for
$\Omega^2$ produced by the conformal transformation).  This is the price 
that is paid for having an unusual
Jordan-frame action.

\section{Relevance to Cosmology}
\label{sec:contrancos}

The purpose of describing all of these non-standard gravity theories and 
their conformal transformations
is to help explain dark energy and the coincidence problem.  To do so we 
must identify the scalar field or fields
with dark energy.  The standard scalar field dark energy model (without a 
conformal transformation) is called
Quintessence \cite{cald98}, so we are just assigning the name 
``quintessence'' to all the scalar fields in any
particular model.  In this way we may use the scalar-matter couplings
generated by conformal transformation to produce a constant ratio of matter
to dark energy and thus to solve the coincidence problem.

\section{Matter-Dark Energy Coupling}
\label{sec:contrandmde}

If we try to model the universe with a stress-energy conservation equation 
like
equation (\ref{matcons}) we quickly run into problems due to the fact that 
$\partial_{\nu} \log \Omega^2$
must be small for the entire history of the universe if cosmological 
observations and tests of 
general relativity are to be believed.  Simply put, we know that
baryons, neutrinos, 
and photons have not had cosmologically significant interactions with
other species.  Before 
decoupling, their interactions appear 
to be consistent with the Standard Model of particle
physics, and after decoupling their stress-energy tensors are independently 
conserved to good approximation. If we look at a 
linear scalar-tensor example
with $\frac{1}{2}\frac{d \log \Omega^2(\phi)}{d \phi} = \xi \sqrt{\frac{16 
\pi G}{3}}$ 
the parameter $|\xi|\sqrt{\frac{16 \pi G}{3}}$ must be less than 0.032 to be 
consistent with radar delay 
experiments \cite{reas79,damo90}.

The problem for conformally transformed theories does not anyway
lie with photons and relativistic neutrinos since their stress-energy 
tensors are trace-free; 
the problem instead lies with cold dark matter, baryons, and 
non-relativistic neutrinos. 
We can state this problem as follows: how do we partition

\begin{equation}
\nabla^{\mu}\left(T_{\mu \nu}^{B}+T_{\mu \nu}^{C}+T_{\mu \nu}^{\nu,nr}
\right) =  -\frac{1}{2}
\Omega^{-2} \partial_{\nu} \Omega^2 \left( T^{B}+T^{C}+T^{\nu,nr}\right)
\end{equation}

\noindent
such that the resulting cosmological model is distinct from the standard 
one and yet is consistent 
with the observational bounds on couplings to visible matter?

One possibility is for each species to couple only to itself:
\begin{eqnarray}
\nabla^{\mu} T_{\mu \nu}^{C} &=& -\frac{1}{2} \Omega^{-2} \partial_{\nu} 
\Omega^2  T^{C} \nonumber\\
\nabla^{\mu} T_{\mu \nu}^{B} &=& -\frac{1}{2} \Omega^{-2} \partial_{\nu} 
\Omega^2  T^{B} \nonumber\\
\nabla^{\mu} T_{\mu \nu}^{\nu,nr} &=& -\frac{1}{2} \Omega^{-2} \partial_{
\nu} \Omega^2 T^{\nu,nr}.
\end{eqnarray}

\noindent
However, since the constraints from the visible sector on $\Omega^{-2} 
\partial_{\nu} \Omega^2$ 
also constrain the dark sector this model is not observationally distinct 
from the 
standard cosmology.  An oft-used alternative due to Damour \textit{et al.}
\cite{damo90} is to postulate that 
 $\Omega^{-2} \partial_{\nu} \Omega^2$ from the latter two equations is 
different from that of the first.  This allows coupling to the dark
sector to be less constrained than coupling to the visible sector.  It is on 
this utilitarian basis that the 
modification (usually with $\Omega^{-2} \partial_{\nu} \Omega^2 = 0$ in the 
visible sector) is used 
in cosmology as the Coupled Quintessence Model \cite{amen00b}. 

It is important to stress that the coupling asymmetry has important 
effects; physically it means that
dark matter and visible matter naturally inhabit two different gravitational 
theories, where the ``natural'' 
theory for a given matter species is here taken to mean the one in which its 
stress-energy tensor is conserved.
Thus the equivalence principle is violated in this model, a not 
inconsiderable loss.  We are also left with a
deeper question of how we interpret the conformal transformation that gave 
rise to these matter couplings.
What is the relationship between the Jordan and Einstein frame theories now?  
Breaking the equivalence principle
actually puts the two frames on more footing since each has its own 
``natural'' matter fields.  From the 
point of view of conserving matter stress-energy they are equally 
valid.  Thus, it seems that the mathematical 
inconsistency results in an appealing interpretation in that 
both frames can be said to be physically relevant.

Another possibility is to have cold dark matter couple to all three species, 
leaving the baryons and 
non-relativistic neutrinos to be exactly conserved:

\begin{eqnarray}\label{cataquint}
\nabla^{\mu} T_{\mu \nu}^{C} &=& -\frac{1}{2}
\Omega^{-2} \partial_{\nu} \Omega^2 \left( T^{B}+T^{C}+T^{\nu,nr}\right) 
\nonumber\\
\nabla^{\mu} T_{\mu \nu}^{B} &=& 0 \nonumber\\
\nabla^{\mu} T_{\mu \nu}^{\nu,nr} &=& 0.
\end{eqnarray}

\noindent
Here the evolution of the visible components is affected by the dark sector
only through the Friedmann equation, but the interaction between dark matter 
and dark energy is 
catalyzed by the presence of baryons and non-relativistic neutrinos.  
Let us refer to this as the Catalyzed Quintessence model.  There is nothing 
mathematically inconsistent with this possibility though it may well be 
physically questionable as it allows
the dark matter energy density to go negative in the Einstein frame, 
though total energy is still conserved.
Although this is certainly not a desirable feature, the model may be
considered as a sort of matter analogue to phantom energy \cite{cald02}.  
The application of this model to cosmology is discussed further 
in \cite{crooks05}.

What is the status of the two frames in this model?  Oddly, the stress-energy 
tensor of visible matter is conserved 
in both frames.  This does not mean that all observations will be the same in 
both frames since the influence of gravity
will in general be quite different, and furthermore dark matter will be 
conserved in one frame and not in the other. 
Thus there is no simple answer to the question, ``Which frame is the physical 
one?''  The criterion of stress-energy
conservation may not give an unambiguous answer.  Additionally, there is no 
symmetry here between the two frames, and indeed the
Jordan frame is arguably the more natural one since all types of matter are
conserved in it.  Thus the natural
frame and the observable frame need not be the same; we have to define the 
observable frame to be physical \textit{a posteriori}.

\section{Asymptotic Behavior: Single Scalar Field}
\label{sec:cosdynasymsingle}

Since we are trying to explain the coincidence problem it suffices to look 
for theories 
whose dynamics contain a stable and accelerated critical point where the 
ratio of dark matter to dark energy is constant.
Thus far the theories presented have generally 
contained arbitrary 
functions, usually in the form of the Jordan frame potential, that must be 
chosen before their dynamics
can be determined.  There are infinitely many possible choices of potential, 
but it turns out that it is
not necessary to search them all to find one with a mixed 
critical point; instead we can work backwards from the 
requirement of a mixed critical point and the equations of motion to 
construct the form that works.  The requirement of acceleration can then be
imposed.  The requirement of stability will not be discussed here as it
depends on the specific parameter values.

For a linear scalar-tensor or nonlinear gravity system in an expanding 
Friedmann-Robertson-Walker (FRW) universe
the scalar field equation of motion in the Einstein frame is  

\begin{equation}\label{singscaleom}
\omega \left( \ddot{\phi} + 3 \frac{\dot{a}}{a} \dot{\phi} \right) + \frac{d 
U(\phi)}{d \phi} = \Upsilon(\phi)
\rho_{M},
\end{equation}

\noindent
while the dark matter energy conservation equation (\ref{regboxeqn}) is

\begin{equation}\label{dmeom}
\dot{\rho}_{M}+3\frac{\dot{a}}{a}\rho_{M} = -\Upsilon(\phi)\rho_{M}\dot{\phi},
\end{equation}

\noindent
where $\Upsilon(\phi) \equiv \frac{1}{2} \Omega^{-2}(\phi) \frac{d \Omega^2 (
\phi)}{d \phi}$.

On the critical point we require the potential, kinetic, and dark matter
energy densities to be exactly proportional:

\begin{equation}
\rho_{M} = \epsilon U(\phi) = \kappa \frac{\omega}{2} \dot{\phi}^2,
\end{equation}

\noindent
where $\epsilon$ and $\kappa$ are positive constants.  Using the Friedmann 
equation on the attractor,

\begin{equation}
\left( \frac{\dot{a}}{a}\right)^2 = \frac{8 \pi G}{3} \left( \frac{\omega}{2}
\dot{\phi}^2 + V(\phi) + 
\rho_{M} \right),
\end{equation}

\noindent 
the conservation equation for $\rho_{M}$, the equation of motion for
$\phi$, and the above proportionality constraints it is possible to show 
that the coupled energy densities all evolve with the scale factor like

\begin{equation}
\rho \sim a^{-\frac{3 \epsilon (\kappa + 2)}{\epsilon + \kappa + 
\epsilon\kappa}}.
\end{equation}

\noindent
Specifically, $\rho_{M}$ has the solution

\begin{equation}
\rho_{M} = \left(\frac{a}{a_{0}}\right)^{-\frac{3 \epsilon (\kappa + 2)}
{\epsilon + \kappa + 
\epsilon\kappa}} \rho_{M,0},
\end{equation}

\noindent
where subscript $M$ denotes either dark matter alone or the combination of 
all matter species depending on which model
is chosen.  On the other hand, the scale factor evolves with time like

\begin{equation}
a(t) = a_{0} \left( \frac{t}{t_{0}} \right)^{\frac{2(\epsilon + \kappa + 
\epsilon \kappa)}
{3 \epsilon (2+\kappa)}}
\end{equation}

\noindent
so that at late time the coupled energy densities evolve like $\rho 
\sim t^{-2}$.

\noindent
From this we may determine the time evolution of $\phi$, which turns out 
to be

\begin{equation}
\phi = \phi_{0} + \frac{1}{\sqrt{ \pi G \omega}}\ln \left[ \left( \frac{t}
{t_{0}} 
\right)^{ \pm \frac{\sqrt{\epsilon + \kappa + \epsilon \kappa}}{(2 + \kappa) 
\sqrt{3 \epsilon}}} \right].
\end{equation}

With the time evolution and the coupling ratios it is then possible to 
calculate the exact form 
that the potential must have for there to be a mixed attractor:

\begin{equation}
U(\phi) = \frac{\rho_{M,0}}{\epsilon} \: e^{\pm (2+\kappa)\sqrt{\frac{12 
\omega \epsilon \pi G}
{\epsilon + \kappa + \epsilon \kappa}}(\phi-\phi_{0}) }, 
\end{equation}

\noindent
where

\begin{equation}
\rho_{M,0}=\frac{\kappa (\epsilon + \kappa + \epsilon \kappa)}{6 \pi G 
\epsilon (2 + \kappa)^2 t_{0}^2}.
\end{equation}

\noindent
If we let $U(\phi) \sim e^{\mu \sqrt{\frac{16 \pi G}{3}} \phi}$ and compare 
this to the required form for 
a critical point to exist we find that the model parameter $\mu$ is related 
to the energy density ratios by

\begin{equation}
\frac{\mu}{\sqrt{\omega}} = \pm \frac{3\sqrt{\epsilon}(2+\kappa)}{2\sqrt{
\epsilon+\kappa+\epsilon\kappa}}.
\end{equation}

\noindent
Furthermore, if we look at the simplest model of interest, that for which 
$\Upsilon(\phi) = \xi 
\sqrt{\frac{16 \pi G}{3}}$, we can use our knowledge of $\rho_{M}(t)$ and 
$\phi(t)$ to simplify the dark
matter conservation equation.  The latter becomes

\begin{equation}\label{singlescalarconstraint}
\frac{\kappa \left(\epsilon+\kappa+\epsilon\kappa\right) \left(3\kappa
-3\epsilon 
\mp 2 \frac{\xi}{\sqrt{\omega}}\sqrt{\epsilon(\epsilon+\kappa+\epsilon\kappa)}
\right)}{6 \epsilon^2 (2+\kappa)} = 0.
\end{equation}

\noindent
Since we are assuming $\epsilon$ and $\gamma$ are positive, this constraint 
gives us the following relation between
the model parameter $\xi$ and the coupling ratios:

\begin{equation}\label{xiconstraint}
\frac{\xi}{\sqrt{\omega}} = \mp \frac{3 (\epsilon-\kappa)}{2\sqrt{\epsilon(
\epsilon+\kappa+\epsilon\kappa)}}.
\end{equation}

Therefore, given $\omega$, $\mu$, and $\xi$ we can write down exactly what 
the various energy densities
(written here as fractions of the critical density) will be at the mixed 
critical point:

\begin{eqnarray}\label{omegasmuk}
\Omega_{M} &=& \frac{2 \mu^2 + 2 \mu \xi - 9 \omega}{2 (\mu + \xi)^2} 
\nonumber\\ 
\Omega_{U} &=& \frac{4 \mu \xi + 4 \xi^2 + 9 \omega}{4 (\mu + \xi)^2} 
\nonumber\\
\Omega_{K} &=& \frac{9 \omega}{4 (\mu + \xi)^2}.
\end{eqnarray}

Looking ahead, it is possible to find the region of the parameter space (the 
model parameters) 
in which the critical point is accelerated by determining where the overall 
equation of state,

\begin{equation}
w = \frac{\Omega_{K}-\Omega_{U}}{\Omega_{K}+\Omega_{U}+\Omega_{M}} = \frac{
\epsilon-\kappa}{\epsilon+\kappa+\epsilon\kappa},
\end{equation}

\noindent
is less than $-\frac{1}{3}$.  Such a region does in fact exist as was shown 
in \cite{crooks05}.

\section{Asymptotic Behavior: Dual Scalar Fields}
\label{sec:cosdynasymdouble}

The quadratic scalar-tensor case is somewhat more complicated.  In an 
expanding homogeneous isotropic universe the
equations (\ref{boxpsi}) and (\ref{boxphi}) take the form : 

\begin{eqnarray}
\omega \left( \ddot{\phi}+3\frac{\dot{a}}{a}\dot{\phi}\right)&=&-\frac{
\partial U(\psi,\phi)}
{\partial \phi}+\sqrt{\frac{4 \pi G \omega}{3}}\left(\rho_{M}-2 
e^{-\sqrt{\frac{16 \pi G \omega}{3}}\phi}\zeta A(\psi)\dot{\psi}^2\right)
\nonumber\\
\ddot{\psi}+3\frac{\dot{a}}{a}\dot{\psi}&=&-e^{\sqrt{\frac{16 \pi G \omega}
{3}}\phi}
\frac{\zeta}{2 A(\psi)}\frac{\partial U(\psi,\phi)}{\partial \psi}
\nonumber\\ & & -\frac{1}{2}
\frac{d \log A(\psi)}{d \psi} \dot{\psi}^2 + 2\sqrt{\frac{4 \pi G \omega}{3}}
\dot{\phi}\dot{\psi},
\end{eqnarray}

\noindent
while equation (\ref{matcons}) reduces to

\begin{equation}
\dot{\rho}_{M} + 3\frac{\dot{a}}{a}\rho_{M} = -\sqrt{\frac{4 \pi G \omega}
{3}}\rho_{M} \dot{\phi}.
\end{equation}

\noindent
Unfortunately, it is not possible to determine what $A(\psi)$ must be to 
allow a mixed 
attractor.  The root of this problem lies in the fact that $A(\psi)$ always 
appears in the combination 
$\zeta A(\psi) \dot{\psi}^2$ or its derivative, which means there are not 
enough equations to uniquely
determine the time evolution of both $A(\psi)$ and $\dot{\psi}$.  We may 
circumvent this problem, 
however, by a change of variables:

\begin{equation}\label{chidefinition}
\chi \equiv \int \sqrt{2 \zeta A(\psi)} d\psi.
\end{equation}

\noindent
The equations of motion become:

\begin{eqnarray}
\omega \left(\ddot{\phi} + 3\frac{\dot{a}}{a}\dot{\phi}\right) + \frac{
\partial U(\chi,\phi)}
{\partial \phi} = \sqrt{\frac{4 \pi G \omega}{3}}\left(\rho_{M}-
e^{-\sqrt{\frac{16 \pi G \omega}{3}}\phi} \dot{\chi}^2\right) \nonumber\\
\ddot{\chi}+3\frac{\dot{a}}{a}\dot{\chi} + e^{\sqrt{\frac{16 \pi G \omega}{3}
}\phi} 
\frac{\partial U(\chi,\phi)}{\partial \chi} = - \sqrt{\frac{16 \pi G \omega}
{3}}\dot{\chi}\dot{\phi},
\end{eqnarray} 

\noindent
which must obey the constraints:

\begin{equation}\label{ekgconstraint}
\rho_{M} = \epsilon U(\chi,\phi) = \kappa \frac{1}{2} e^{-\sqrt{\frac{16 
\pi G}{3}} \phi} 
\dot{\chi}^2 = \gamma \frac{\omega}{2} \dot{\phi}^2.
\end{equation}

The total energy density evolves with the scale factor like

\begin{equation}
\rho \sim a^{-\frac{3 \epsilon (2 \kappa + 2 \gamma + \kappa \gamma)}{
\kappa \gamma + 
\gamma \epsilon + \epsilon \kappa + \gamma \epsilon \kappa}},
\end{equation}

\noindent
while the scale factor in turn evolves with time as

\begin{equation}
a = a_{0} \left( \frac{t}{t_{0}} \right)^{\frac{2 (\kappa \gamma + \gamma 
\epsilon 
+ \epsilon \kappa + \gamma \epsilon \kappa)}{3 \epsilon (2 \kappa + 
2 \gamma + \kappa \gamma)}}.
\end{equation}

Thus the total energy density evolves with time as $\rho \sim t^{-2}$ just 
like in the single
scalar case.  Also analogous to the single scalar case, the field $\phi$ 
in the two-field case
evolves with time as

\begin{equation}\label{phiofteqn}
\phi = \phi_{0} + \frac{1}{ \sqrt{\pi G \omega}}\ln \left[ \left( \frac{t}
{t_{0}} 
\right)^{\pm \frac{1}{2 \mu}} \right],
\end{equation}

\noindent
whereas the new field $\chi$ evolves as

\begin{equation}
\chi = \chi_{0} \left( \frac{t}{t_{0}} \right)^{\pm \frac{1}{\mu}}.
\end{equation}

\noindent
Here $\mu$ is a dimensionless ratio defined as:

\begin{equation}
\mu = \frac{3 \sqrt{\epsilon} (2 \kappa + 2 \gamma 
+ \kappa \gamma)}{2 \sqrt{\kappa (\kappa \gamma 
+ \gamma \epsilon + \epsilon \kappa + \gamma \epsilon \kappa)}},
\end{equation}

\noindent
which is again similar to the single scalar case.  Using this time 
dependence of $\phi$ in the dark matter
conservation equation we get the constraint

\begin{equation}\label{epkagaconstraint}
\frac{\gamma\kappa \left(\epsilon\kappa+\kappa\gamma+\epsilon\gamma
+\epsilon\gamma\kappa\right) 
\left(3\epsilon\gamma+3\epsilon\kappa-3\kappa\gamma 
\mp \sqrt{\epsilon\kappa(\epsilon\kappa+\kappa\gamma+\epsilon\gamma
+\epsilon\gamma\kappa)}\right)}{9 \epsilon^2 
(2\kappa+2\gamma+\gamma\kappa)} = 0,
\end{equation}

\noindent
which is analogous to equation (\ref{xiconstraint}) relating $\xi$ to 
$\epsilon$ and $\kappa$ in the single 
scalar example.  However, there is no variable $\xi$ in the two-scalar 
case (it is effectively set to $\frac{1}{2}$)
so this equation acts as a constraint on $\epsilon$, $\kappa$, and $\gamma$ 
at the critical point rather 
than relating these variables to a model parameter.  
For the critical point to exist it must satisfy 
this constraint, but in order to be accelerated
it must also have an equation of state satisfying

\begin{equation}\label{wdoublescal} 
w = \frac{\gamma\epsilon+\kappa\epsilon-\kappa\gamma}{\gamma\epsilon
+\kappa\epsilon+\kappa\gamma+\kappa\epsilon\gamma} <
-\frac{1}{3}. 
\end{equation} 

\noindent
Unfortunately, the regions of $\{\kappa,\epsilon,\gamma\}$-space in which
the constraint and the acceleration condition 
are satisfied do not intersect at any point.  Thus, the hybrid model cannot 
both realistically model our universe and 
solve the coincidence problem.

\section{Asymptotic Behavior: General Case}
\label{sec:cosdynasymgeneral}

Is there a deeper reason why the dual scalar model does not work but the 
single scalar model does?  Certainly the failure 
of the former has nothing to do with the changes of variables since the same 
dark matter constraint and condition 
for acceleration result when we use $\psi$ instead of $\chi$ as well as when 
we use $\Omega^2$ instead of $\psi$.  Furthermore,
having more scalar fields (as in a Higher-Order Gravity model) does not help 
matters.  To see this, let the Jordan frame
action have a combined potential $\widetilde{U}(\psi_{0},\dots,\psi_{n-1},
\Omega^2)$ and a combined kinetic term
$\widetilde{g}^{\mu \nu}\widetilde{K}_{\mu \nu}$ for the $\psi_{j}$.  Note 
that we are not assuming any specific form 
for the individual kinetic terms save that they have two lower tensor 
indices.  The individual terms may be multiplied 
by functions of the scalar fields (including $\Omega^2$), and they may even 
contain higher derivatives.  
The Einstein frame action in this example is 

\begin{eqnarray}
S &=& \int d^{4}x \left\{\sqrt{-g} \left[ \frac{R}{16 \pi G}-g^{\mu \nu}K_{
\mu \nu}+
U(\psi_{0},\dots,\psi_{n-1},\Omega^2)\right.\right.\nonumber\\
& &\left.\left.-\frac{3}{32 \pi G}\Omega^{-4} g^{\mu \nu}\partial_{\mu}
\Omega^2 \partial_{\nu}\Omega^2\right]
-L^{(M)}\right\},
\end{eqnarray}

\noindent
where $U(\psi_{0},\dots,\psi_{n-1},\Omega^2)=\Omega^{-4}\widetilde{U}
(\psi_{0},\dots,\psi_{n-1},\Omega^2)$ and 
$K_{\mu \nu} = \Omega^{-2}\widetilde{K}_{\mu \nu}$.  The Einstein frame 
scalar field stress-energy tensor is

\begin{eqnarray}
T_{\mu \nu}(\psi_{0},\dots,\psi_{n-1},\Omega^2) &=& 2 \left(K_{\mu \nu} 
-\frac{1}{2}g_{\mu \nu}g^{\alpha \beta}
K_{\alpha \beta}\right) + g_{\mu \nu} U(\psi_{0},\dots,\psi_{n-1},\Omega^2
)\nonumber\\
&&{} +\frac{3}{32 \pi G}\Omega^{-4}
\left(\partial_{\mu}\Omega^2 \partial_{\nu} \Omega^2 -\frac{1}{2} g_{
\mu \nu} g^{\alpha \beta}\partial_{\alpha}\Omega^2
\partial_{\beta}\Omega^2\right),\nonumber\\&&
\end{eqnarray}

\noindent
and its conservation in an FRW cosmology yields

\begin{eqnarray}\label{conservationofK}
\frac{1}{2}\Omega^{-2}\frac{\partial \Omega^2}{\partial t}\rho_{M} &=& 
\frac{\partial}{\partial t}\left(K_{0 0}
+U(\psi_{0},\dots,\psi_{n-1},\Omega^2)+\frac{3}{32 \pi G}\Omega^{-4}\left(
\frac{\partial \Omega^2}{\partial t}\right)^2 \right)\nonumber\\ &&+ 3\frac{
\dot{a}}{a}\left(2 K_{0 0}+\frac{3}{16 \pi G}
\Omega^{-4}\left(\frac{\partial \Omega^2}{\partial t}\right)^2\right).
\end{eqnarray}

\noindent
Here we have assumed that only the $0-0$ component of $K_{\mu \nu}$ is 
relevant to the homogeneous evolution, the other 
components entering the picture at level of first- or higher-order 
perturbations.  This is an extremely generic assumption
since there is no easy way to combine scalars into an object with two 
tensor indices besides including covariant derivatives,
and only the $0$-component of a covariant derivative contributes to the 
homogeneous evolution of the universe. One
alternative would be to build $K_{\mu \nu}$ from the metric $g_{\mu \nu}$ 
multiplied by some function of the scalars; 
however, such an object could be re-interpreted as a potential term and 
would therefore not belong in $K_{\mu \nu}$ at all.
Another possibility would be to use the four dimensional Levi-Civita density 
$\varepsilon_{\mu \nu \lambda \sigma}$ to mix up 
the indices of two contravariant derivatives, but doing so would always pair 
time derivatives with space derivatives so
that every term would be negligible to zero-th order.  Therefore let us 
assume that $K_{\mu \nu}$ is dominated by its
$0-0$ component.

To be on a critical point each scalar field must have a kinetic term either 
equal to zero or proportional to 
the energy density of dark matter.  This implies that all non-vanishing 
kinetic terms must also be proportional to each other.  
Therefore, without loss of generality, we can treat 
$K_{\mu \nu}$ as a whole and require it to be proportional to the dark
matter energy density.  We now have

\begin{equation}
\rho_{M} = \epsilon U(\psi,\phi) = \kappa K_{0 0} 
= \gamma \frac{3}{32 \pi G}\Omega^{-4}\left(\frac{\partial \Omega^2}
{\partial t}\right)^2,
\end{equation}

\noindent
which is a generalization of equation (\ref{ekgconstraint}).

If we follow a procedure similar to that used in Section 
\ref{sec:cosdynasymdouble} for the dual scalar case we find that together
these ratios along with the Friedmann equation and the equations of motion 
yield the exact same constraints on $\epsilon$, $\kappa$,
and $\gamma$ as were found in equation (\ref{epkagaconstraint}).
Furthermore, if the $0-0$ component of $K_{\mu \nu}$ is the only one 
relevant to the
homogeneous evolution of the universe then the equation of state for this 
general kinetic term is the same as for a
standard kinetic term, \textit{i.e.}, $w=1$. Therefore, the constraint that 
the overall equation of state be less than 
$-\frac{1}{3}$ is the same as the two-scalar case above, equation 
(\ref{wdoublescal}).

Thus having a kinetic term for $\Omega^2$ distinct from other kinetic terms 
for the $\psi_{j}$ prevents
the model from solving the coincidence problem and also allowing 
accelerated expansion.  But then why does the
single scalar case work?  It works because the field $\Omega^2$ 
is just a redefinition of the field $\psi$, 
which allows us to combine the kinetic terms in equation (\ref{einslagr}):

\begin{equation}\label{generalLagrangian}
L^{(\psi)} = \sqrt{-g} \left[\zeta A(\psi) \Omega^{-2}(\psi)+\frac{3}{32 
\pi G} \Omega^{-4} \left( 
\frac{\partial \Omega^2}{\partial \psi} \right)^2\right] 
g^{\mu \nu} \partial_{\mu} \psi \partial_{\nu} \psi -\sqrt{-g} U(\psi).
\end{equation}

\noindent
Let us refer to the quantity inside the brackets as $\frac{1}{2}
\omega(\psi)$.  Previously, we transformed this 
Lagrangian into one having a canonical kinetic term via the redefinition
$\phi = \int d \psi \sqrt{\omega(\psi)}$.
However, whether we choose to make this redefinition or not has no effect on
the present discussion, and it is 
more instructive to use the general form.  The scalar field equation of
motion and the dark matter conservation equation
are:

\begin{eqnarray}
\omega(\psi)\left(\ddot{\psi}+3\frac{\dot{a}}{a}\psi \right)+\frac{1}{2}
\frac{d\omega(\psi)}{d\psi}\dot{\psi}^2+\frac{d U(\psi)}
{d\psi} &=& \Upsilon(\psi) \rho_{M} \nonumber\\
\dot{\rho}_{M}+3\frac{\dot{a}}{a}\rho_{M} &=& -\Upsilon(\psi)\dot{\psi}
\rho_{M},
\end{eqnarray}

\noindent
and the condition for being on a critical point is

\begin{equation}
\rho_{M} = \epsilon U(\psi) = \kappa \frac{\omega(\psi)}{2} \dot{\psi}^2.
\end{equation}

If we then follow the same procedure as before for finding $\rho_{M}(a)$ 
and $a(t)$ and then insert these
into the dark energy conservation equation we find, instead of a constraint 
on the energy ratios or a relation
between these and a model parameter, a more general constraint on 
$\omega(\psi)$ and $\Upsilon(\psi)$:

\begin{equation}
\frac{\kappa \left(\epsilon+\kappa+\epsilon\kappa\right) \left(3\kappa
-3\epsilon 
\mp \frac{\sqrt{3}\Upsilon(\psi)}{\sqrt{4 \pi G \omega(\psi)}}\sqrt{\epsilon(
\epsilon+\kappa+\epsilon\kappa)}\right)}
{6 \epsilon^2 (2+\kappa)} = 0.
\end{equation}

\noindent
If $\kappa$ and $\epsilon+\kappa+\epsilon\kappa$ are non-zero then we have 
that the ratio 

\begin{equation}
\frac{\Upsilon(\psi)}{\sqrt{\frac{4 \pi G}{3} \omega(\psi)}} = \frac{
\frac{1}{2}\Omega^{-2}(\psi)\frac{d \Omega^2 (\psi)}
{d \psi}}{\sqrt{\frac{8 \pi G}{3} \zeta A(\psi) \Omega^{-2}(\psi)+\left(
\frac{1}{2} \Omega^{-2}(\psi) \frac{d \Omega^2(\psi)}
{d \psi}\right)^2}}
\end{equation}

\noindent
must be constant if the universe to is be on a mixed critical point.  A 
comparison with equation
(\ref{singlescalarconstraint}) shows that this constant value is what we 
have been calling 
$2\frac{\xi}{\sqrt{\omega}}$.  The problem with the multiple scalar situation 
is that, since we cannot combine the kinetic
terms for the $\psi_{j}$ and $\Omega^2$, the prefactor on the $\Omega^2$
kinetic term is just 
$\frac{3}{32 \pi G} \Omega^{-4} \left( 
\frac{\partial \Omega^2}{\partial \psi} \right)^2$ instead of the more 
complicated expression in 
equation (\ref{generalLagrangian}).  Thus, in the multiple scalar case we 
find that the constant value 
is actually just $\pm 1$, and, as we shall soon show, there is no way to get 
an accelerated and mixed critical point
with these values.

We have just seen that the condition of constant energy ratios enforces a 
relationship between the functions 
$\omega(\psi)$ and $\Upsilon(\psi)$; what about the other undetermined 
function, $U(\psi)$?  It turns out that constant
energy ratios also fix the form of $U(\psi)$, which is then related to 
$\omega(\psi)$ by

\begin{equation}
U(\psi) = \frac{\rho_{M,0}}{\epsilon}\exp\left\{\mu \sqrt{\frac{16 \pi G}
{3}} 
\int_{\psi_{0}}^{\psi}\sqrt{\omega(\psi')}\;d\psi'\right\}.
\end{equation}

\noindent
Therefore, whenever the lone kinetic term is set to have constant positive 
coupling to gravity (via transformation or careful 
selection of $A(\psi)$ and $\Omega^{2}(\psi)$) the potential must be of 
exponential form in order for a mixed critical point to exist.

\section{Possible Generalizations}
\label{sec:cosdynposgeneral}

In order for a universe exhibiting dark matter-dark energy coupling to 
evolve on a mixed critical point we have just argued that the
potential must be related to the prefactor on the dark energy kinetic term;
specifically, when the dark energy is minimally
coupled we showed that the potential must be of exponential form.  Let us 
assume this to be the case.  Obviously a realistic
universe does not evolve on the mixed critical point for all time; we 
require only an asymptotic approach to the point.  This 
means that the requirement of an exponential potential is too tight.  All 
that is required is an exponential tail that is
energetically accessible from the rest of the potential.  

For example, a potential formed from the sum of two exponentials 
(having parameters $\mu_{1}$ and $\mu_{2}$) will have two mixed, accelerated 
points for each one point in the single
exponential case, \textit{i.e.}, one for each $\mu$.  
If the two $\mu$s have the same sign 
the stable point will correspond to the 
potential term with the lower value of $|\mu|$ since that term will dominate 
on the tail.  
If the $\mu$s have different signs the potential is U-shaped, and therefore 
the only stable point will be 
the one corresponding to the global minimum.

Another simple possibility is the potential derived from compactification of 
the Kaluza-Klein extra dimension (see \cite{ferr98}
for a discussion of this and other examples),

\begin{equation}
U(\psi) = e^{-\psi/\sqrt{8 \pi G}}\left(1-e^{-\psi/\sqrt{8 \pi G}}\right)^2,
\end{equation}

\noindent
which has an exponential tail.  This potential has another local minimum at 
the origin, though, so it has two possible late-time 
behaviors depending on $\dot{\psi}$.  If $\dot{\psi}$ is below a certain 
threshold, the field oscillates about the 
origin, but if $\dot{\psi}$ is large enough, the field climbs the ``hump'' 
onto the exponential tail.

\section{Conclusions}

This article has discussed conformally transformed gravity theories and their 
possible use in solving the Coincidence Problem.  
Four types of gravity theories were examined.  Depending on whether the 
original Jordan frame theory contained scalar fields,
or functions, higher powers, or derivatives of the Ricci scalar, the 
resulting Einstein frame action involved 
scalar fields and a potential that could be completely, partial, or not 
at all determined.  Indeed, 
there was found to be a degeneracy among Jordan frame theories in that there 
can be many with the same 
Einstein frame physics. 

For a given Jordan frame theory there are several possible strategies for 
using a conformal transformation in a realistic
cosmological scenario.  The conformal transformation generates scalar fields 
in the Einstein frame that may be interpreted as 
cosmological quintessence, and it induces a coupling between these 
fields and the matter fields.  There are a number ways of implementing this 
coupling depending on which matter fields 
are coupled, as well as which ones are present in the coupling term.  The 
choice for the former need not imply the same
choice for the latter as the Catalyzed Quintessence Model demonstrates.  
The physical interpretation of the conformal 
transformation itself can depend on these choices.

Many of the more complicated conformally transformed theories, \textit{i.e.}, 
those theories which
contain more than one scalar field in the Einstein frame, were unable to 
solve the Coincidence Problem.  
Furthermore, for a single scalar field model
to solve the Coincidence Problem there are constraints on the relationships 
between the scalar field coupling to gravity 
$\omega(\psi)$, the dark matter-dark energy coupling $\Upsilon(\psi)$, and 
the scalar potential $U(\psi)$ which must be obeyed.
First, the quantity $\Upsilon(\psi)/\sqrt{\omega(\psi)}$ must equal some 
non-trivial constant or at least
asymptotically approach one; second, the potential must have the form 
$U(\psi) \sim \exp\left\{
\mu \sqrt{\frac{16 \pi G}{3}} \int_{\psi_{0}}^{\psi}\sqrt{\omega(\psi')}
d\psi'\right\}$ though perhaps only on the tail.

\section{Acknowledgements}

We thank H. Karwowski, C. Clemens, Y.J. Ng, H. van Dam, 
J. Engel, and K. Crooks for their comments. 
This work was supported in part by the
U.S. Department of Energy under Grant
No. DE-FG02-97ER-41036

\end{document}